\documentstyle[11pt]{article}
\catcode`\@=11
 
\@addtoreset{equation}{section}
\@addtoreset{equation}{subsection}
\def\theequation{\ifnum\value{subsection}>0\relax
\thesubsection.\arabic{equation}\relax
\else\ifnum\value{section}>0\relax
\thesection.\arabic{equation}\relax
\else\arabic{equation}\fi\fi}

\newcommand{\cR}{{\cal R}}

\newcommand{\bK}{{\bf K}}
\newcommand{\sn}{\;{\rm sn}\,}
\newcommand{\cn}{\;{\rm cn}\,}
\newcommand{\dn}{\;{\rm dn}\,}
\newcommand{\half}{{1\over{2}}}
\def\sl#1{\not{\hbox{\kern-2pt ${#1}$}}}

\def\th{\theta}

\begin{document}

\title{\Large \bf Supersymmetric Sine-Gordon Model and the Eight-Vertex
Free Fermion model with Boundary }

\author{Changrim Ahn\\
\small Department of Physics\\
\small Ewha Womans University\\
\small Seoul 120-750, Korea\\ \\
and \\ \\
Wai Ming Koo\\
\small Center for Theoretical Physics\\
\small Seoul National University\\
\small Seoul 151-742, Korea}
\date{May 1996}

\begin{titlepage}
\maketitle

\begin{abstract}

We compute the boundary scattering amplitudes 
of the breathers of the supersymmetric sine-Gordon model using 
the fusion of the soliton-antisoliton pair
scattering with the boundary 
with a known result of the soliton boundary scattering amplitudes.
We also solve the boundary Yang-Baxter equation of the 
eight-vertex free fermion models to find the boundary reflection matrices.
The former result is confirmed by the latter 
since the bulk $S$-matrices of the 
breathers can be identified with the trigonometric limit of the
Boltzmann weights of the free fermion models.
Our dual approach can answer a few quesions on the relationships between the
free parameters in the boundary potential and those in the scattering
amplitudes.

\end{abstract}
\vspace{2cm}
\rightline{SNUTP-96-046}
\rightline{EWHA-TH-009}
\vspace{1cm}
\end{titlepage}

\section{Introduction and Motivation}

The study of the two-dimensional integrable models of quantum field
theories and statistical models based on the Yang-Baxter equation (YBE)
has provided important understandings
of nonperturbative aspects of these models and technical tools
for applications to real physical problems.
The YBE plays essential roles in establishing the integrability and
solving the models. In the field theories, the YBE provides a consistency
condition for the two-body scattering amplitudes ($S$-matrices) in the
multi-particle scattering processes since the scattering is factorizable.
With unitarity and crossing symmetry, the YBE can determine the $S$-matrix
completely.
In addition, the correlation functions may be obtained by computing
multi-particle form-factors.
The lattice models which are defined by the Boltzmann weights
can have well-defined transfer matrices if they satisfy the YBE
and can be diagonalized by independent technologies,
such as the algebraic Bethe ansatz to extract exact properties of the model.

Recently there has been a lot of efforts in extending these
approaches to models with boundaries.
They are motivated by the fact that
these models with the boundary have more applicability to 
real physical systems than those without one.
For example, three-dimensional spherically symmetric physical
systems can be effectively described on the half-line
if $s$-wave element becomes dominant.
One-channel Kondo problem, monopole-catalyzed proton decay are
frequently cited examples.
Also one can generalize the conventional periodic boundary condition
of the statistical models to other types like the fixed and free conditions.

The existence of the boundary adds new quantities like
boundary scattering amplitudes and Boltzmann weights,
and one needs to extend the YBE to include these objects.
The boundary Yang-Baxter equation (BYBE) (also known as the reflection
equation) \cite{cher} plays the role of the YBE for the integrable statistical
models \cite{skly,oth} and quantum field theories \cite{zam} in the
presence of a boundary; it is the necessary condition for the
integrability of these models.

The supersymmetric sine-Gordon model (SSG) preserves the integrability in the
presence of the boundary \cite{inam} and the scattering amplitude of the
SSG solitons with it has been computed \cite{ssg}.
In this paper we compute the boundary scattering amplitudes of 
the SSG breathers, the bound states of the soliton and antisoliton pair, 
in two independent ways.
The first one is to consider the the eight-vertex free fermion models
with the boundary and to solve the BYBE.
This is related to the SSG model since the breather $S$-matrices 
are trigonometric limit of the free fermion models in certain regimes.
The second is to compute the two-particle 
(a SSG soliton and an antisoliton) boundary scattering amplitudes
and to take a limit of two rapidities so that they can form a bound state.  
This `fusion procedure' can give independent check of our results.
In addition, these two approaches can answer a few questions raised 
in the previous study;
how the supersymmetry of the SSG lagrangian with the boundary can
be realized in the scattering matrix context and
how the parameters in the lagrangian and those in
the boundary scattering amplitudes are related.

\section{SSG Breathers on a half line}

The action of the SSG model is given by
\begin{equation}
S= \int dx dt \left[ \half(\partial_{\mu}\phi)^2
-i{\overline\psi}\sl{\partial}\psi - {m^2\over{\beta^2}}\cos^2\phi
- 2m(\cos{\beta\phi\over{2}}){\overline\psi}\psi \right],\label{eq:action}
\end{equation}
where $\phi$ is a real scalar field and  $\psi$ is a Majorana fermion.
$\beta$ is a coupling constant and $m$ is the mass
parameter denoting the deviation from the massless theory.
This theory has soliton spectrum $\vert K^{\pm}_{ab}(\theta)\rangle$
where `$ab$' and `$\pm$' are the RSOS spins ($a,b=0,\half,1$) and the 
topological charges (`$+$' for the soliton and `$-$' for the antisoliton),
respectively and $\theta$ is the rapidity.
The exact $S$-matrix of the SSG (anti-)solitons has
the factorized form of \cite{bl}
\[
S_{\rm SSG}(\theta)=S^{(4)}_{\rm RSG} (\theta)\otimes S_{\rm SG}(\theta).
\]
The first $S$-matrix factor which acts on the supersymmetry (SUSY) 
charges is the RSOS $S$-matrix for the tricritical Ising model 
perturbed by the $\Phi_{13}$ operator;
\begin{equation}
S^{(4)}_{\rm RSG}=
S^{ab}_{dc}(\th)=U(\th)\left(X^{ab}_{cd}\right)^{-{\th\over{2\pi i}}}
\left[\sqrt{X^{ab}_{cd}}\sinh\left({\th\over{p}}\right)\delta_{db}+
\sinh\left({i\pi-\th\over{p}}\right)\delta_{ac}\right],
\end{equation}
for $\vert\bK_{da}(\th_1)\rangle+\vert\bK_{ab}(\th_2)
\rangle\to\vert\bK_{dc}(\th_2)\rangle +\vert\bK_{cb}(\th_1)\rangle$
where
$X^{ab}_{cd}=\left({[2a+1][2c+1]\over{[2d+1][2b+1]}}\right)$ with
$q$-number $[n]=(q^{n}-q^{-n})/(q-q^{-1})$ and $q=-e^{-i\pi/4}$ (Fig.1).
\vskip .5cm
\begin{picture}(400,100)(0,0)
\thinlines
\put (100,0) {\line(1,1){100}}
\put (200,0) {\line(-1,1){100}}
\put (115,50) {$d$}
\put (150,85) {$a$}
\put (185,50) {$b$}
\put (150,15) {$c$}
\end{picture}
\vskip .5cm
\noindent
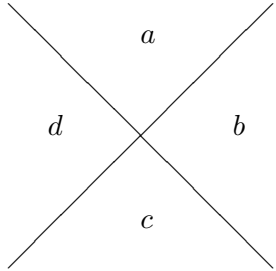
\centerline{{\bf Fig. 1}  Bulk $S$-matrix of the kinks}
\vskip 1cm
The second one is formally the sine-Gordon (SG) (anti)soliton $S$-matrix 
with the parameter given by $\gamma=4\beta^{2}/(1-\beta^2/4\pi)$.\footnote{ 
Notice that the $S$-matrix of the SG model depends on 
$\gamma=\beta^{2}/(1-\beta^2/8\pi)$.}
The factorized form of the $S$-matrix implies that the SSG soliton 
can be formally written as $\vert\bK^{\pm}_{ab}(\theta)\rangle
=\vert\bK_{ab}(\theta)\rangle\otimes\vert\pm(\theta)\rangle$.
The charge conjugation of the SSG solitons is defined by 
\begin{equation}
C\vert\bK^{\pm}_{ab}\rangle=\vert\bK^{\mp}_{ba}\rangle.\label{eq:charge} 
\end{equation}
For $n<8\pi/\gamma\le n+1$, 
the second factor, the SG $S$-matrix, has $n$ poles in the physical strip
corresponding to the SSG breathers. 
The threshold value of the SSG $\beta$ to have any bound
state is $\beta^{2}=4\pi/3$ compared with that of the SG, $\beta^{2}=4\pi$.

The bulk $S$-matrices of the breathers have been obtained by considering 
the residues of two solitons and two antisolitons scattering and taking
appropriate limits on the rapidities.
Due to the factorized form of the soliton $S$-matrix, 
the SSG breather $S$-matrices are also made up of two factors. 
The factors coming from the SG sector have been computed in \cite{karow} 
and they are completely diagonal
since the masses of the SG breathers are non-degenerate.
The second one comes from the four kinks scattering fusion processes
of the RSG(4) \cite{ahn}.
It is obvious that this factor is non-diagonal since the SUSY
makes the mass spectrum degenerate and the breathers form $N=1$
supermultiplets.

Since two kinks can scatter only when they share a common RSOS spin, 
the two-kink states which form the SSG breathers can be written as
\begin{equation}
\begin{array}{lll} 
|\psi_n^1(\theta)\rangle&=&\frac{\textstyle i\alpha_n}{\textstyle\sqrt{2}}
\left(|\bK_{0\frac{1}{2}}
(\theta_1)\bK_{\frac{1}{2}1}(\theta_2)\rangle-|\bK_{1\frac{1}{2}}(\theta_1)
\bK_{\frac{1}{2}0}(\theta_2)\rangle\right)\;,\vspace{3mm}\\
|\phi_n^1(\theta)\rangle&=&\frac{\textstyle 1}{\textstyle \sqrt{2}}
\left(|\bK_{0\frac{1}{2}}(\theta_1)
\bK_{\frac{1}{2}0}(\theta_2)\rangle+|\bK_{1\frac{1}{2}}(\theta_1)
\bK_{\frac{1}{2}1}(\theta_2)\rangle\right)\;,\vspace{3mm}\\
|\psi_n^2(\theta)\rangle&=&\frac{\textstyle \alpha_n}{\textstyle\sqrt{2}}
\left(|\bK_{\frac{1}{2}0}
(\theta_1)\bK_{0\frac{1}{2}}(\theta_2)\rangle-|\bK_{\frac{1}{2}1}(\theta_1)
\bK_{1\frac{1}{2}}(\theta_2)\rangle\right)\;,\vspace{3mm}\\
|\phi_n^2(\theta)\rangle&=&\frac{\textstyle 1}{\textstyle \sqrt{2}}
\left(|\bK_{\frac{1}{2}0}(\theta_1)
\bK_{0\frac{1}{2}}(\theta_2)\rangle+|\bK_{\frac{1}{2}1}(\theta_1)
\bK_{1\frac{1}{2}}(\theta_2)\rangle\right)
\end{array}\label{eq:fusion}
\end{equation}
where the rapidities are related as
\[\theta=\frac{1}{2}\left(\theta_1+\theta_2\right)\;,\quad\theta_1-\theta_2
=\triangle\theta_n\]
and 
\[\alpha_n=\sqrt{\tan\left(\frac{\pi+i\triangle\theta_n}{4}\right)}\;,\quad
\triangle\theta_n=i\pi-\frac{in\gamma}{8}\;,\]
with $\triangle\theta_n$ corresponding to the mass pole of the breathers.

Notice that the bound states come in two sets distinguished by their
superscripts $1,2$ and only particles with the same superscripts can scatter. 
Each pair $(\psi_n^a,\phi_n^a)$ forms $N=1$ supermultiplet 
(see Eq.(\ref{eq:susy})) and has the same bulk $S$-matrices, 
\cite{ahn}
\begin{equation}
S(\theta)=\rho(\theta)\left(\begin{array}{cccc}
\sin\frac{n\gamma}{16}+\frac{i}{2}\sinh \theta&&&-i\sin\frac{n\gamma}{16}
\sinh\frac{\theta}{2}\\
&-\frac{i}{2}\sinh \theta&\sin\frac{n\gamma}{16}\cosh\frac{\theta}{2}&\\
&\sin\frac{n\gamma}{16}\cosh\frac{\theta}{2}&-\frac{i}{2}\sinh \theta&\\
-i\sin\frac{n\gamma}{16}\sinh\frac{\theta}{2}&&&\sin\frac{n\gamma}{16}-
\frac{i}{2}\sinh \theta\;
\end{array}\right),\label{eq:sw}
\end{equation}
where the column and row are arranged in the order of 
$\psi_n^a\psi_n^a$, $\psi_n^a\phi_n^a$, $\phi_n^a\psi_n^a$, 
$\phi_n^a\phi_n^a$ for $a=1,2$.
The states in Eq.(\ref{eq:fusion}) are invariant under charge conjugation $C$, 
implying that they are real scalar particles and Majorana fermions. 
The least massive bound states ($n=1$) are identified with 
the $\psi$ and $\phi$ fields in the Lagrangian Eq.(\ref{eq:action})
and, indeed, the above $S$-matrix is identical to that obtained 
in \cite{shan} if we identify $\sin\frac{\gamma}{16}$ with $f$. 
Since there is only one fundamental field pair $(\psi,\phi)$ in
the lagrangian, the two sets of the bound states which have the 
same $S$-matrices should be indentified,
i.e. $|\psi_n^1(\theta)\rangle \equiv|\psi_n^2(\theta)\rangle$ 
and $|\phi_n^1(\theta)\rangle\equiv |\phi_n^2(\theta)\rangle$. 

The function $\rho(\theta)$ satisfies the unitarity and crossing relations
\begin{equation}
\begin{array}{l}
\rho(\theta)\rho(-\theta)(\sin^2\frac{n\gamma}{16}+\sinh^2\frac{\theta}{2})
\cosh^2\frac{\theta}{2}=1\vspace{3mm}\\
\rho(\theta)=\rho(i\pi-\theta)
\end{array}\;.
\end{equation}
The minimum solution to these equations have been given as \cite{shan}
\begin{equation}
\rho(\theta)=-\frac{2i}{\sinh\theta}Z(\theta)Z(i\pi-\theta)
\end{equation}
where
\begin{eqnarray*}
Z(\theta)&=&\frac{\textstyle \Gamma(-i\theta/2\pi)}
{\textstyle \Gamma(1/2-i\theta/2\pi)}
\prod_{l=1}^{\infty}\left[\frac{\textstyle
\Gamma(n\gamma/16\pi-i\theta/2\pi+l)\Gamma(-n\gamma/16\pi-i\theta/2\pi+l-1)}
{\textstyle \Gamma(n\gamma/16\pi-i\theta/2\pi+l+1/2)
\Gamma(-n\gamma/16\pi-i\theta/2\pi+l-1/2)}\right.\\
&&\mbox{}\times\left.\frac{\textstyle \Gamma^2(-i\theta/2\pi+l-1/2)}{\textstyle
 \Gamma^2(-i\theta/2\pi+l-1)}\right]\;.
\end{eqnarray*}

Now we introduce a boundary potential which preserves the integrability.
The SSG boundary potential that gives conserved charges 
at the first order has been derived as \cite{inam}:
\begin{equation}
{\cal B}(\phi,\psi,{\overline\psi})=\Lambda\cos{\beta(\phi-\phi_0)\over{2}}
+M{\overline\psi}\psi+\epsilon\psi+{\overline\epsilon}{\overline\psi}.
\label{eq:inami}
\end{equation}
With the assumption of the complete integrability, one can use the BYBE
to solve this model.
For the purpose, we use
an important property of the $S$-matrix in Eq.(\ref{eq:sw}) that 
it satisfies the free fermion condition \cite{fan,ahn}. 
Therefore it should be a special limit of the Boltzmann weights 
of the eight-vertex free fermion model given in the Appendix. 
Indeed, consider the regime $|h|>1$ and
take the following trigonometric limit:
\[k\rightarrow 0\quad\mbox{ with }\quad k\cosh\delta=k\sinh\delta\equiv 
-{1\over{\sin{n\gamma\over{16}}}}\;.\]
We obtain the breather $S$-matrix if $u\equiv-i\theta$.
An immediate consequence of this observation is that 
the boundary $S$-matrix of the SSG breathers are given by
the trigonometric limit of the boundary Boltzmann weights
derived in the Appendix.
The trigonometric limit of Eq.(\ref{eq:eswb}) becomes
\begin{equation}
R(\theta)=\cR(\theta)\left(\begin{array}{cc}
\cosh\frac{\theta}{2}G_{+}(\theta)-i\sinh\frac{\theta}{2}G_{-}(\theta)&
-i\epsilon\sinh \theta\vspace{3mm}\\
-i\sinh \theta&\cosh\frac{\theta}{2}G_{+}(\theta)+i\sinh\frac{\theta}{2}
G_{-}(\theta) \end{array}\right)\label{eq:swb}
\end{equation}
where 
\begin{equation}
G_{\pm}(\theta)=\alpha_{\pm}-\frac{\epsilon\alpha_{\pm}+\alpha_{\mp}}
{\sin{n\gamma\over{16}}-\epsilon}\sinh^2\frac{\theta}{2}\label{eq:alpmi}
\end{equation}
and
\begin{equation}
\alpha_{+}^2-\alpha_{-}^2=2\left(\epsilon-{1\over{\sin{n\gamma\over{16}}}}
\right)\;,\quad\quad \epsilon=\pm 1\label{eq:rel}\;
\end{equation}
The overall factor $\rho(\theta), \cR(\theta)$ are functions that
ensure the unitarity and crossing symmetry of the $S$- and $R$- matrices,
respectively.

The equations that determine $\cR(\theta)$ are given by
\begin{equation}
\begin{array}{l}
\cR(\theta)\cR(-\theta)(\cosh^2\frac{\theta}{2}G_{+}^2(\theta)+
\sinh^2\frac{\theta}{2}G_{-}^2(\theta)+\epsilon\sinh^2\theta)=1\vspace{3mm}\\
\rho(2\theta)\cR(\frac{i\pi}{2}+\theta)\cosh\theta\left(\sin\frac{n\gamma}{16}
-i\epsilon\sinh\theta\right)=\cR(\frac{i\pi}{2}-\theta)\;.
\end{array}
\end{equation}
As usual we define
\[\cR(\theta)=\cR_0(\theta)\cR_1(\theta)\]
such that 
\begin{equation}
\begin{array}{l}
\cR_1(\theta)=\cR_1(i\pi-\theta)\vspace{3mm}\\
\cR_1(\theta)\cR_1(-\theta)\left(c_0+c_1\sinh^2\frac{\theta}{2}+c_2\sinh^4\frac{\theta}{4}\right)=1
\end{array}
\end{equation}
where 
\[c_0=\alpha^2_{+}\;,\quad c_1=\frac{\epsilon\left(\alpha_{+}
+\epsilon\alpha_{-}\right)^2}{\epsilon-\sin\frac{n\gamma}{16}}+2\epsilon\;,
\quad c_2=\frac{\left(\alpha_{+}+\epsilon\alpha_{-}\right)^2}
{\left(\epsilon-\sin\frac{n\gamma}{16}\right)^2}\;,\]
and
\begin{equation}
\begin{array}{l}
\rho(2\theta)\cR_0(\frac{i\pi}{2}+\theta)\cosh\theta\left(
\sin\frac{n\gamma}{16}-i\epsilon\sinh\theta\right)=\cR_0(\frac{i\pi}{2}-
\theta)\vspace{3mm}\\
\cR_0(\theta)\cR_0(-\theta)\cosh\theta=1.
\end{array}\;
\end{equation}
The factor $\cR_1(\theta)$ carries information about the
boundary conditions that are determined by the free parameters $\alpha_{\pm}$,
and its minimum solution is given by
\begin{equation}
\cR_1(\theta)=\frac{1}{\alpha_{+}}\sigma(\chi,\theta)\sigma(\eta,\theta)
\end{equation}
where the function $\sigma(\chi,\theta)$ is an infinite product of
$\Gamma$ function defined as
\begin{eqnarray*}
\sigma(\chi,\theta)&=&\frac{\textstyle \Pi(\chi,\pi/2+i\theta)
\Pi(-\chi,\pi/2+i\theta)\Pi(\chi,-\pi/2-i\theta)\Pi(\chi,-\pi/2-i\theta)}
{\textstyle \Pi^2(\chi,\pi/2)\Pi^2(-\chi,\pi/2)}\;,\\
\Pi(\chi,-i\theta)&=&\prod_{l=1}^{\infty}\frac{\textstyle \Gamma(l+\chi/\pi
+i\theta/2\pi)}{\textstyle \Gamma(l+1+\chi/\pi+i\theta/2\pi)}\;,
\end{eqnarray*}
with the parameters $\chi,\eta$ defined by
\[\cos^{-2}\chi+\cos^{-2}\eta=c_1/c_0\;,\quad \cos^{-2}\chi\cos^{-2}\eta
=c_2/c_0\;.\]
The relations that determine $\cR_0(\theta)$ can, similarly, 
be solved with minimum solution given by
\begin{equation}
\cR_0(\theta)=\frac{Y(\theta)Y(i\pi-\theta)}{\pi
\sqrt{i\epsilon\sinh(2\theta)\ \rho(-\pi/2-2i\theta)}}
\end{equation}
where
\[Y(\theta)=\prod_{l=1}^{\infty}\frac{\textstyle 
\Gamma(1-l+\epsilon n\gamma/36\pi+1/4+i\theta/2\pi) 
\Gamma(l-\epsilon n\gamma/36\pi-1/4-i\theta/2\pi)} 
{\textstyle \Gamma(-l+\epsilon n\gamma/36\pi+1/4-i\theta/2\pi) 
\Gamma(l+1-\epsilon n\gamma/36\pi-1/4+i\theta/2\pi)}\;. \]

\newpage
\section{Soliton Fusion}

In \cite{ssg}, we studied the scattering theory of the SSG model 
on a half line based on its soliton states. 
Essentially, in the presence of a boundary, 
integrability of the SSG model requires that the boundary $S$-matrix of 
the solitons satisfy the BYBE, Eq.(\ref{eq:bybe}). 
We can assume naturally that this boundary
$S$-matrix is also factorized into two parts:
\begin{equation}
R_{\rm SSG}(\theta)=R_{RSG}^{(4)}(\theta)\otimes R_{\rm SG}(\theta)
\end{equation}
where $R_{\rm SG}$ and $R_{RSG}^{(4)}$ are the SG
and  RSOS(4) boundary scattering matrices, respectively. 
Writing it in this form, we can solve the BYBE separately.
The SG part has been obtained in \cite{zam,dev} and 
the kinks part has been found in \cite{ssg} to be of the form (Fig.2)
\begin{equation}  
R^{a}_{bc}(\theta)=\cR(\theta)\left(X^{bc}_{aa}
\right)^{-\frac{\theta}{2\pi i}} \left[\delta_{b\neq c}X^{a}_{bc}
(\theta)+\delta_{bc}\left(\delta_{b-1/2,a} U_{a}(\theta)
+\delta_{b+1/2,a}D_{a}(\theta)\right)\right]
\end{equation}
where $a,b,c$ are the RSOS(4) spins.
\vskip -.5cm
\begin{picture}(400,100)(0,0)
\thicklines
\put (100,0) {\line(1,0){150}}
\thinlines
\put (175,.7) {\line(1,1){50}}
\put (175,.7) {\line(-1,1){50}}
\put (125,20) {$b$}
\put (225,20) {$a$}
\put (175,40) {$c$}
\end{picture}
\vskip .5cm
\noindent
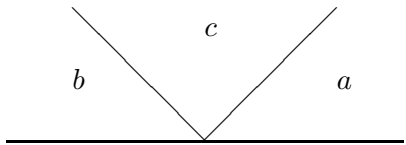
\centerline{{\bf Fig.2} Boundary kink $S$-matrix}
\vskip 1cm
The explicit solutions are given by
\begin{equation}\begin{array}{l}
X^{\frac{1}{2}}_{01}=sX^{\frac{1}{2}}_{10}\;,\quad 
\frac{\textstyle U_{\frac{1}{2}}(\theta)}{\textstyle 
X_{01}^{\frac{1}{2}}}=\frac{\textstyle B}
{\textstyle \sinh\frac{\theta}{2}}
+C\cosh\frac{\theta}{2}\;,\quad \frac{\textstyle D_{\frac{1}{2}}(\theta)}
{\textstyle X_{01}^{\frac{1}{2}}}=
\frac{\textstyle B}{\textstyle\sinh\frac{\theta}{2}}
-C\cosh\frac{\theta}{2}\;,\vspace{3mm}\\
\frac{\textstyle D_{1}(\theta)}{\textstyle U_0(\theta)}=
\frac{\textstyle 1-A\sinh\frac{\theta}{2}}
{\textstyle 1+A\sinh\frac{\theta}{2}}
\end{array}\label{eq:bsm}
\end{equation}
with $A,B,C$ being the free parameters of the boundary, and
the off-diagonal terms $X^{\frac{1}{2}}_{01},X^{\frac{1}{2}}_{10}$ are 
independent
of the spectral parameter and differ from each other by a gauge
factor $s$. The overall function $\cR(\theta)$ that guarantees boundary
crossing and unitarity is given in \cite{ssg}.

As we have shown that the bound states of the RSG(4) kinks give rise to
the $\psi_n(\theta),\phi_{n}(\theta)$ fields, whose bulk scattering
matrix is given by Eq.(\ref{eq:sw}), the scattering of these 
$\psi_n,\phi_n$ fields with the boundary is governed by the matrix
given in Eq.(\ref{eq:swb}). Hence we expect that by fusing the boundary 
scattering matrices of the kinks given above, we should reproduce  
Eq.(\ref{eq:swb}). Before proceeding with the computation, it is
worth recalling that, the fermionic and bosonic bound states come
in two types; $\left(\psi^{1}_{n},\phi^{1}_{n}\right),
\left(\psi^{2}_{n},\phi^{2}_{n}\right)$, which are, nevertheless, identified
in the bulk since the scattering
matrices of these two sets of particles have exactly the same form and
are hence indistinguishable from each other. It would be natural to
wonder whether the same holds true in the presence of a boundary. In fact,
more intriguingly, notice that the boundary scattering matrices of the
kinks carry more than one free parameters, while the scattering matrix
given in Eq.(\ref{eq:swb}) has only one. From the fusion equation,
it is clear that the $R$-matrix of $\left(\psi^2_n,\phi^2_n\right)$, which 
will be built out of $U_0, D_1$, will contain one free parameter as in 
Eq.(\ref{eq:swb}). However, that of $\left(\psi^1_n,\phi^1_n\right)$, 
which will be built from $X^{\frac{1}{2}}_{01}$, $X^{\frac{1}{2}}_{10}$,
$U_{\frac{1}{2}}$, and $D_{\frac{1}{2}}$, 
will contain more than one free parameter, and this will be incompatible with
Eq.(\ref{eq:swb}) as they have the same bulk $S$-matrix given by 
Eq.(\ref{eq:sw}). Also interesting is to try to clarify the
relation of the two classes of solution, distinguished by $\epsilon=\pm 1$,
with the fused boundary $S$-matrix of the kinks. 

We represent the boundary scattering 
$|\bK_{ab}(\theta_1)\bK_{bc}(\theta_2)\rangle\longrightarrow
|\bK_{cd}(-\theta_1)\bK_{de}(-\theta_2)\rangle$
in Fig.3.
\vskip .5cm
\begin{picture}(400,100)(0,0)
\thicklines
\put (50,0) {\line(1,0){240}}
\thinlines
\put (130,.7) {\line(1,1){90}}
\put (130,.7) {\line(-1,1){50}}
\put (210,.7) {\line(1,1){50}}
\put (210,.7) {\line(-1,1){90}}
\put  (90,10) {$a$}
\put (130,40) {$b$}
\put (170,80) {$c$}
\put (170,10) {$f$}
\put (210,40) {$d$}
\put (250,10) {$e$}
\end{picture}
\vskip .5cm
\noindent
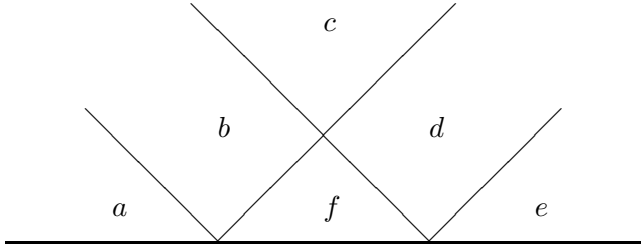
\centerline{{\bf Fig.3} Boudnary scattering of the two-kink states}
\vskip 1cm
Let us begin with the bound states $\left(\psi^2_n,\phi^2_n\right)$.
Using the fusion equations Eq.(\ref{eq:fusion}), we can construct their 
boundary $S$-matrix by combining $U_{0},D_{1}$ and the bulk $S$-matrix 
of the kinks as follows:
\begin{equation}
\begin{array}{lll}
R^2_{\psi,\psi}(\theta)&=&\frac{1}{2}\left(U_0(\theta_1)U_0(\theta_2)
S^{0\frac{1}{2}}_{\frac{1}{2}0}(\theta_1+\theta_2)-U_0(\theta_1)D_1(\theta_2)
S^{0\frac{1}{2}}_{\frac{1}{2}1}(\theta_1+\theta_2)\right.\\
&&\left.+D_1(\theta_1)D_1(\theta_2)
S^{1\frac{1}{2}}_{\frac{1}{2}1}(\theta_1+\theta_2)-D_1(\theta_1)U_0(\theta_2)
S^{1\frac{1}{2}}_{\frac{1}{2}0}(\theta_1+\theta_2)\right)\vspace{3mm}\\
R^2_{\psi,\phi}(\theta)&=&\frac{1}{2\alpha_n}\left(U_0(\theta_1)U_0(\theta_2)
S^{0\frac{1}{2}}_{\frac{1}{2}0}(\theta_1+\theta_2)+D_1(\theta_1)U_0(\theta_2)
S^{1\frac{1}{2}}_{\frac{1}{2}0}(\theta_1+\theta_2)\right.\\
&&\left.-U_0(\theta_1)D_1(\theta_2)
S^{0\frac{1}{2}}_{\frac{1}{2}1}(\theta_1+\theta_2)-D_1(\theta_1)D_1(\theta_2)
S^{1\frac{1}{2}}_{\frac{1}{2}1}(\theta_1+\theta_2)\right)\vspace{3mm}\\
R^2_{\phi,\psi}(\theta)&=&\frac{\alpha_n}{2}\left(U_0(\theta_1)U_0(\theta_2)
S^{0\frac{1}{2}}_{\frac{1}{2}0}(\theta_1+\theta_2)+U_0(\theta_1)D_1(\theta_2)
S^{0\frac{1}{2}}_{\frac{1}{2}1}(\theta_1+\theta_2)\right.\\
&&\left.-D_1(\theta_1)U_0(\theta_2)
S^{1\frac{1}{2}}_{\frac{1}{2}0}(\theta_1+\theta_2)-D_1(\theta_1)D_1(\theta_2)
S^{1\frac{1}{2}}_{\frac{1}{2}1}(\theta_1+\theta_2)\right)\vspace{3mm}\\
R^2_{\phi,\phi}(\theta)&=&\frac{1}{2}\left(U_0(\theta_1)U_0(\theta_2)
S^{0\frac{1}{2}}_{\frac{1}{2}0}(\theta_1+\theta_2)+U_0(\theta_1)D_1(\theta_2)
S^{0\frac{1}{2}}_{\frac{1}{2}1}(\theta_1+\theta_2)\right.\\
&&\left.+D_1(\theta_1)U_0(\theta_2)
S^{1\frac{1}{2}}_{\frac{1}{2}0}(\theta_1+\theta_2)+D_1(\theta_1)D_1(\theta_2)
S^{1\frac{1}{2}}_{\frac{1}{2}1}(\theta_1+\theta_2)\right)
\end{array}
\end{equation}
where the superscript 2 on the boundary $S$-matrix refers to the
second set of bound states. 

As an example the explicit computations for 
$R^2_{\psi, \psi}(\theta)$ and $R^2_{\phi,\psi}$ look as follows:
\begin{eqnarray*}
R^2_{\psi,\psi}(\theta)&=&\frac{U_0(\theta_1)U_0(\theta_2)}
{\sqrt{2}\left(1+A\sinh\frac{\theta_1}{2}\right)
\left(1+A\sinh\frac{\theta_2}{2}\right)}\left[\cosh\frac{\theta}{2}
\left(1-A^2\sinh^2\frac{\triangle\theta_n}{4}+A^2\sinh^2\frac{\theta}{2}
\right)\right.\\
&&\mbox{}\left.-i\sinh\frac{\theta}{2}
\left(-1-A^2\sinh^2\frac{\triangle\theta_n}{4}+A^2\sinh^2\frac{\theta}{2}
\right) \right]\;,\\
R^2_{\phi,\psi}&=&\frac{i\alpha_1 A\sinh\theta U_0(\theta_1)U_0(\theta_2)}
{\sqrt{2}\left(1+A\sinh\frac{\theta_1}{2}\right)
\left(1+A\sinh\frac{\theta_2}{2}\right)}
\left(\cosh\frac{\triangle\theta_n}{4}-i\sinh\frac{\triangle\theta_n}{4}
\right)\;.
\end{eqnarray*}

Comparing them with the elements in the first column of the 
boundary $S$-matrix given in Eq.(\ref{eq:swb}), and after dividing by 
an overall factor
\[-\frac{\alpha_1 A U_0(\theta_1)U_0(\theta_2)}
{\sqrt{2}\left(1+A\sinh\frac{\theta_1}{2}\right)
\left(1+A\sinh\frac{\theta_2}{2}\right)}\left(\cosh\frac{\triangle
\theta_n}{4}-i\sinh\frac{\triangle\theta_n}{4}\right)\;,\]
we see that the above $R^2_{\psi,\psi}$ has the form
\begin{equation}\cosh\frac{\theta}{2}G_{+}(\theta)-i\sinh\frac{\theta}{2}G_{-}
(\theta)
\end{equation}
with 
\begin{equation}
G_{\pm}(\theta)=-\left(\alpha_1A\right)^{-1}\left(\cosh
\frac{\triangle\theta_n}{4}-i\sinh\frac{\triangle\theta_n}{4}\right)^{-1}
\left(\pm 1-A^2\sinh^2\frac{\triangle\theta_n}{4}+A^2\sinh^2\frac{\theta}{2}
\right)\;.\label{eq:alpmii}
\end{equation}

Comparing Eq.(\ref{eq:alpmii}) with Eq.(\ref{eq:alpmi}), we
deduce that this fused $S$-matrix corresponds to the $\epsilon=1$ case
since the coefficients of $\sinh^2\frac{\theta}{2}$ are the same. 
Moreover, considering $\theta$-independent terms in both equations, we find 
\begin{equation}
\alpha_{\pm}=-\left(\alpha_1A\right)^{-1}\left(\cosh
\frac{\triangle\theta_n}{4}-i\sinh\frac{\triangle\theta_n}{4}\right)^{-1}
\left(\pm 1-A^2\sinh^2\frac{\triangle\theta_n}{4}\right)\;.
\end{equation} 
Notice that the coefficient of the $\sinh^2\frac{\theta}{2}$-term in 
Eq.(\ref{eq:alpmii}) is given by
\[
-\frac{\alpha_{+}+\alpha_{-}}{\sin{n\gamma\over{16}}-1}.
\]
Substituting the expressions for $\alpha_{\pm}$ into this, we can 
produce the corresponding coefficient in Eq.(\ref{eq:alpmi}).
In addition the $\alpha_{\pm}$ satisfy 
\[\alpha^2_{+}-\alpha_{-}^2=2\left(1-{1\over{\sin{n\gamma\over{16}}}}
\right)\;,\]
which is consistent with Eq.(\ref{eq:rel}).
The same computations for the other two components 
$R^2_{\psi,\phi},R^2_{\phi,\phi}$ also lead to the same conclusion as above. 
We therefore confirm that this fused boundary $S$-matrix indeed reproduces 
Eq.(\ref{eq:swb}) with $\epsilon=1$ and the mapping of the
boundary parameter is given by
\begin{equation}
iA=\frac{\alpha_{+}+\alpha_{-}}{1-\left(\sin{n\gamma\over{16}}\right)^{-1}}\;.
\end{equation}

When $\left(\psi^1_n,\phi_n^1\right)$ scatter with the boundary,
the out-states are linear combinations of $\left(\psi^1_n,\phi_n^1\right)$
and two other states.
For example we have
\begin{eqnarray}
R(\theta)|\phi^1_n(\theta)\rangle&=&R^1_{\phi,\phi}(\theta)
|\phi^1_n(-\theta)\rangle+R^1_{\psi,\phi}(\theta)
|\psi^1_n(-\theta)\rangle\nonumber\\
&&\mbox{}+BC\sinh\theta\cosh\frac{\triangle\theta_n}{2}
\left(\cosh\frac{\theta}{2}-i\sinh\frac{\theta}{2}\right)|\Omega_1(-\theta)
\rangle\nonumber\\
&&\mbox{}-\frac{1}{2}\sinh\theta
\left(\cosh\frac{\triangle\theta_n}{4}+i\sinh\frac{\triangle\theta_n}{4}
\right)\nonumber\\
&&\mbox{}\times\left[B\left(1+s\right)
+i\frac{C}{2}\left(1-s\right)
\left(\cosh\theta-\cosh\triangle\theta_n\right)\right]
|\Omega_2(-\theta)\rangle\;,
\end{eqnarray}
where $R^1_{\phi,\phi}, R^1_{\psi,\phi}$ are amplitudes to be given later.
The two new states are given by
\begin{eqnarray*}
|\Omega_1(\theta)\rangle&\equiv&
\frac{1}{\sqrt{2}}\left(|\bK_{0\frac{1}{2}}(\theta_1)\bK_{\frac{1}{2}0}
(\theta_2)\rangle-
|\bK_{1\frac{1}{2}}(\theta_1)\bK_{\frac{1}{2}1}(\theta_2)\rangle\right)\;,\\
|\Omega_2(\theta)\rangle&\equiv&\frac{1}{\sqrt{2}}\left(
|\bK_{0\frac{1}{2}}(\theta_1)\bK_{\frac{1}{2}1}(\theta_2)\rangle+
|\bK_{1\frac{1}{2}}(\theta_1)\bK_{\frac{1}{2}0}(\theta_2)\rangle\right)
\end{eqnarray*}
which are orthogonal to $\left(\psi^1_n,\phi_n^1\right)$ states.

These new states do not scatter with 
$\left(\psi^1_n,\phi_n^1\right)$ at all in the bulk and their bulk $S$-matrix
is different from Eq.(\ref{eq:sw}). 
Moreover, the boundary potential in Eq.(\ref{eq:inami}) contains only 
the $(\psi,\phi)$ pair and we should not expect any new particle
to be created by the action of the boundary.
Therefore, we should eliminate these extra states appearing in the 
fusion procedure so that the out-states are the linear 
combinations of only $\left(\psi^1_n,\phi_n^1\right)$. 
This is possible if we choose 
\begin{equation}
B=0 \quad \mbox{ and }\quad s=1\quad\quad \mbox{ or }\quad\quad
C=0 \quad \mbox{ and }\quad  s=-1\;.
\end{equation}
The same conclusion is also arrived when the scattering of
$|\psi^1_n(\theta)\rangle$ with the boundary is considered.

With the above restriction we can reproduce Eq.(\ref{eq:swb})
by fusing the boundary and bulk $S$-matrices of the kinks as follows:
\begin{equation}
\begin{array}{lll}
R^1_{\psi,\psi}(\theta)&=&\frac{1}{2}\left(U_{\frac{1}{2}}(\theta_1)
U_{\frac{1}{2}}(\theta_2)S^{\frac{1}{2}0}_{1\frac{1}{2}}(\theta_1+\theta_2)
+D_{\frac{1}{2}}(\theta_1)D_{\frac{1}{2}}(\theta_2)
S^{\frac{1}{2}1}_{0\frac{1}{2}}(\theta_1+\theta_2)\right.\\
&&\left.+s S^{\frac{1}{2}1}_{1\frac{1}{2}}(\theta_1+\theta_2)+s
S^{\frac{1}{2}0}_{0\frac{1}{2}}(\theta_1+\theta_2)\right)\vspace{3mm}\\
R^1_{\psi,\phi}(\theta)&=&\frac{-i}{2\alpha_n}\left(sD_{\frac{1}{2}}(\theta_2)
S^{\frac{1}{2}1}_{0\frac{1}{2}}(\theta_1+\theta_2)+sU_{\frac{1}{2}}(\theta_1)
S^{\frac{1}{2}1}_{1\frac{1}{2}}(\theta_1+\theta_2)\right.\\
&&\left.-D_{\frac{1}{2}}(\theta_1)
S^{\frac{1}{2}0}_{0\frac{1}{2}}(\theta_1+\theta_2)-U_{\frac{1}{2}}(\theta_2)
S^{\frac{1}{2}0}_{1\frac{1}{2}}(\theta_1+\theta_2)\right)\vspace{3mm}\\
R^1_{\phi,\psi}(\theta)&=&\frac{i\alpha_n}{2}\left(D_{\frac{1}{2}}(\theta_1)
S^{\frac{1}{2}1}_{0\frac{1}{2}}(\theta_1+\theta_2)+U_{\frac{1}{2}}(\theta_2)
S^{\frac{1}{2}1}_{1\frac{1}{2}}(\theta_1+\theta_2)\right.\\
&&\left.-sD_{\frac{1}{2}}(\theta_2)
S^{\frac{1}{2}0}_{0\frac{1}{2}}(\theta_1+\theta_2)-sU_{\frac{1}{2}}(\theta_1)
S^{\frac{1}{2}0}_{1\frac{1}{2}}(\theta_1+\theta_2)\right)\vspace{3mm}\\
R^1_{\phi,\phi}(\theta)&=&\frac{1}{2}\left(U_{\frac{1}{2}}(\theta_1)
U_{\frac{1}{2}}(\theta_2)S^{\frac{1}{2}1}_{1\frac{1}{2}}(\theta_1+\theta_2)
+D_{\frac{1}{2}}(\theta_1)D_{\frac{1}{2}}(\theta_2)
S^{\frac{1}{2}0}_{0\frac{1}{2}}(\theta_1+\theta_2)\right.\\
&&\left.+sS^{\frac{1}{2}0}_{1\frac{1}{2}}(\theta_1+\theta_2)+sS^{\frac{1}{2}1}_{0\frac{1}{2}}
(\theta_1+\theta_2)\right)\;.
\end{array}
\end{equation}
Repeating the analysis given for $R^2_{\psi,\psi}$ before on the
above amplitudes for the two cases, we can show that these
scattering amplitudes coincide with that obtained in Eq.(\ref{eq:swb}). 
We can summarize these in the following mappings:
\begin{equation}
\begin{array}{rllllll}
\mbox{either}&
B=0&s=1&C=\frac{\textstyle -i(\alpha_{+}-\alpha_{-})}{\textstyle 
\left(\sin{n\gamma\over{16}}\right)^{\frac{1}{2}}-
\left(\sin{n\gamma\over{16}}\right)^{-\frac{1}{2}}}
&\epsilon=-1\vspace{3mm}\\
\mbox{or}&C=0&s=-1&B=\frac{\textstyle 
\left(\sin{n\gamma\over{16}}\right)^{\frac{1}{2}}-
\left(\sin{n\gamma\over{16}}\right)^{-\frac{1}{2}}}{\textstyle 
\alpha_{+}+\alpha_{-}}&\epsilon=1\;.
\end{array}
\end{equation}
The two classes ($\epsilon=\pm 1$) of the solutions presented 
in Eq.(\ref{eq:swb}) are indeed compatible 
with the boundary $S$-matrices obtained from the bound states. 

Another aspect of the result is that the two $N=1$ supermultiplets
distinguished by the superscripts $1,2$ scatter
differently with the boundary since the scattering of 
$\left(\psi^1_n,\phi_n^1\right)$ is given by the $S$-matrices with 
$\epsilon=\pm 1$, while that of $\left(\psi^2_n,\phi_n^2\right)$  
by the $S$-matrix with $\epsilon=1$. 
Again we can require that the boundary potential should not add any new
particle states in the theory and the identification of 
$|\psi^1_n\rangle\equiv |\psi^2_n\rangle$ and 
$|\phi^1_n\rangle\equiv |\phi^2_n\rangle$ made in the bulk 
should hold for the boundary as well. 
This dictates that the boundary scattering amplitudes for these two
multiplets be identical. 
Thus we can find that  
\[
C=0, s=1,\quad {\rm and}\quad iAB=\sqrt{\sin{n\gamma\over{16}}}
\]
with $\epsilon=1$, and there 
is only one free parameter associated with the boundary. 

\section{Boundary Supersymmetry}

The SSG model has a supersymmetry and the bound states $(\psi_n, \phi_n)$
(we will drop the superscripts from now on)
transform into each other under the action
of the SUSY charges $Q, \overline{Q}$ in the following way \cite{ahn}:
\begin{equation}
\begin{array}{lllllll}
Q|\phi_n(\theta)\rangle&=&\sqrt{im_n}{\rm e}^{\theta/2}
|\psi_n(\theta)\rangle\;,&\quad\quad&
\overline{Q}|\phi_n(\theta)\rangle&=&\sqrt{-im_n}{\rm e}^{-\theta/2}
|\psi_n(\theta)\rangle\;,\vspace{2mm}\\
Q|\psi_n(\theta)\rangle&=&\sqrt{-im_n}{\rm e}^{\theta/2}
|\phi_n(\theta)\rangle\;,&\quad\quad&
\overline{Q}|\psi_n(\theta)\rangle&=&\sqrt{im_n}{\rm e}^{-\theta/2}
|\phi_n(\theta)\rangle\;,
\end{array} \label{eq:susy}
\end{equation}
where $m_n=2\sin{n\gamma\over{16}}$ is the mass of the $n$-th bound state 
and we have not included the bound states of the SG sector
since the SUSY charges act trivially on them.

One can show that the bulk $S$-matrix of the bound states is invariant
under the action of the SUSY charges, namely,
\begin{equation}
S_{12}(\theta)Q_{12}(\theta)=Q_{21}(-\theta)S_{12}(\theta)\;,\quad\quad 
S_{12}(\theta)\overline{Q}_{12}(\theta)=
\overline{Q}_{21}(-\theta)S_{12}(\theta)\;.
\end{equation}
It is interesting to examine the action of these SUSY charges on the boundary
$S$-matrix to see whether SUSY can be maintained with the
presence of the boundary. 
In fact, it is argued 
that one can retain only `half' of the supersymmetry $Q\pm\overline{Q}$ 
in the presence of a boundary \cite{warner}. 
We shall see that this is the case for the SSG model.
 
One can write the SUSY charges $Q$ and $\overline{Q}$ as $2\times 2$ 
matrices when acting on one-particle state as 
\begin{equation}
\begin{array}{lll}
Q&=&\left(\begin{array}{ll}
          0&{\rm e}^{\frac{\theta}{2}-\frac{i\pi}{4}}\\
          {\rm e}^{\frac{\theta}{2}+\frac{i\pi}{4}}&0
\end{array}\right)\;,\\
\overline{Q}&=&\left(\begin{array}{ll}
          0&{\rm e}^{-\frac{\theta}{2}+\frac{i\pi}{4}}\\
          {\rm e}^{-\frac{\theta}{2}-\frac{i\pi}{4}}&0
\end{array}\right)
\end{array}
\end{equation}
where we have arranged the basis in the order of $\psi_n,\phi_n$.

Consider the action of the linear combination of charges 
$Q(\theta)+\beta\overline{Q}(\theta)$ on the boundary $S$-matrix:
\[R(\theta)\left[Q(\theta)+\beta\overline{Q}(\theta)\right]-
\left[Q(-\theta)+\beta\overline{Q}(-\theta)\right]R(\theta)\;.\]
Using Eq.(\ref{eq:swb}), we deduce that for the above to vanish,
we have either 
\begin{equation}
\beta=1\quad\mbox{ for }\quad \alpha_{+}+\alpha_{-}=0
\end{equation}
or
\begin{equation}
\beta=-1\quad\mbox{ for }\quad \alpha_{+}-\alpha_{-}=0\;.
\end{equation}
From Eq.(\ref{eq:rel}) we see that for the relation $\alpha_{+}=\pm \alpha_{-}$
to be true, we must have $|\alpha_{\pm}|\rightarrow\infty$. In this
limit, the boundary $S$-matrix becomes diagonal
\begin{equation}
R(\theta)=\cR_0(\theta)\left(\begin{array}{cc}
\cosh\frac{\theta}{2}+i\beta\sinh\frac{\theta}{2}&0 \vspace{3mm}\\
0&\cosh\frac{\theta}{2}-i\beta\sinh\frac{\theta}{2}
 \end{array}\right)\label{eq:swbd}
\end{equation}
and there is no boundary free
parameter left, a conclusion which is in agreement with \cite{inam} that
uses a different approach.

We would like to comment that this SUSY preserving boundary $S$-matrix,
when regarded as the boundary reflection matrix of a lattice model, is the
one that gives rise to a spin chain that possesses
the quantum group $U_{1,q^2}gl(1|1)$ symmetry.

\section{Discussion}

In this paper we considered the SSG model with boundary from two points of
view. One is to consider the BYBE of the SSG breathers which are 
related to the eight-vertex free fermion model and the other is to 
use fusion of two kinks which are related to RSOS(4) model.
By matching the boundary $S$-matrices obtained from these two
approaches and requiring that there is only one $N=1$ supermultiplet 
for the bound states, we reduced the four parameters 
$s, A, B, C$ in the SUSY sector of the SSG soliton scattering 
with the boundary to one, and also constrained the solutions of 
the BYBE to the free fermion model to that with $\epsilon=1$.

If we change $\beta$ to $i\beta$ in the SSG model, we get the supersymmetric
sinh-Gordon model with the difference that 
$(\psi_1,\phi_1)$ are the only particles which are not soliton-antisoliton 
bound states as there are no solitons (anti-solitons) in the theory.
Therefore the above argument for the sine-Gordon theory does not 
apply here and both of the $\epsilon=\pm 1$ solutions are allowed.
Since these solutions are mutually exclusive and associated with 
boundary potentials, they must be two different classes of the boundary
potential which preserve integrability.
In addition, each of these potentials should have at least one boundary
coupling parameter which is related to that in the boundary $S$-matrix.
 
The SSG boundary potential in Eq.\ref{eq:inami}, derived from the 
condition that there exist the 
conserved charges at the first order, contains five parameters. 
If we compare this with the SSG boundary soliton $S$-matrix, which has only 
three parameters (two from the sine-Gordon \cite{ghos} and one from the
SUSY sector), we can think of two possibilities;
One is that two of the five parameters in Eq.(\ref{eq:inami}) should
disappear when higher order conserved charges are constructed.
The other possibility is that all the five parameters survive and 
the soliton boundary $S$-matrix introduce two more parameters in its
overall CDD factor.

Furthermore, boundary SUSY is realized only when the five parameters
are fixed as; 
\begin{equation}
\Lambda=\pm 8, M=\pm 1, \phi_{0}=0, \epsilon={\overline\epsilon}=0.
\end{equation}
So there is no free parameters and the $\pm$ are related to the 
``half" SUSY $Q\pm\overline{Q}$ respectively.
For the boundary $S$-matrix, we found that indeed only ``half" SUSY
can be realized. In which case, the boundary $S$-matrix becomes 
diagonal given in Eq.(\ref{eq:swbd}) without any parameters, but it also 
comes in two classes which possess the ``half" SUSY respectively.
This means there is no off-diagonal ($\phi\to\psi,\psi\to\phi$) scattering
amplitude and can be understood from the fact that
the boundary potential deos not includes any fermion number violating term 
as far as $\epsilon={\overline\epsilon}=0$. 
This SUSY, however, acts trivially on the sine-Gordon soliton sector, 
which contains two free parameters, this leads to the conclusion 
that at the SUSY points 
these two parameters become unphysical.

\section*{Acknowledgment}
CA is supported in part by KOSEF 95-0701-04-01-3, 961-0201-006-2  
and BSRI 95-2427 and WMK by a grant from KOSEF through CTP/SNU. 

\section*{Note Added}
After we completed and submitted our paper, we have noticed
that \cite{schout} obtained a similar result as ours on the 
boundary supersymmetry covered in sect.4. We thank the authors
for the information. 

\newpage
\section*{Appendix:}

\centerline{\Large\bf The Eight-Vertex Free Fermion Model with Boundary}
\vskip .5cm
We present solutions to the boundary Yang-Baxter 
equation for the general eight-vertex free fermion model here.

Recall the boundary Yang-Baxter equation takes the form \cite{cher,skly}
\begin{equation}
K_1(u_1)R_{12}(u_1+u_2)K_2(u_2)R_{21}(u_2-u_1)=R_{12}(u_2-u_1)K_2(u_2)
R_{21}(u_1+u_2)K_1(u_1) \label{eq:bybe}
\end{equation}
where $R(u)$ and $K(u)$ are, respectively, the bulk and boundary $R$-matrices.

The eight-vertex free fermion model has been studied by a number of 
authors \cite{fan,fel,baz,cue,rit,gom} the bulk $R$-matrix takes the form
\begin{equation}
R(u)=\left(\begin{array}{llll}
a_{+}&&&d\\
&b_{+}&c&\\
&c&b_{-}&\\
d&&&a_{-}\end{array}
\right)
\end{equation}
where $a_{\pm},b_{\pm},c,d$ denote the usual vertex weights that depend on
the spectral parameter $u$ and other parameters of the model. These weights
satisfy the free fermion condition
\begin{equation}
a_{+}a_{-}+b_{+}b_{-}-c^2-d^2=0
\end{equation}
and $R$-matrices with the same parameters $\Gamma$ and $h$ given
by
\begin{equation}
\Gamma=\frac{2cd}{a_{+}b_{-}+a_{-}b_{+}}\quad\quad\quad h=\frac{a_{-}^2+
b_{+}^2-a_{+}^2-b_{-}^2 }{2\left(a_{+}b_{-}+a_{-}b_{+}\right)}
\end{equation}
commute. 
 
The extreme anisotropic limit of the bulk $R$-matrix
commutes with the quantum spin chain with local Hamiltonian given by
\begin{equation}
{\cal H}_{j,j+1}=\sigma^{+}_{j}\sigma^{-}_{j+1}+\sigma^{-}_{j}
\sigma^{+}_{j+1}+\Gamma\left(\sigma^{+}_{j}\sigma^{+}_{j}+\sigma^{-}_{j}
\sigma^{-}_{j+1}\right)-\frac{h}{2}\left(\sigma^{z}_{j}+\sigma_{j+1}^{z}
\right)\;.
\end{equation}

Thus $h$ can be interpreted as the bulk magnetic field of the
spin chain. Using duality transformation or other argument, one can 
show that critical points must occur for $|h|=1$. Here we shall 
adopt the parameterization given in \cite{fel}; For $|h|<1$ the
vertex weights are given by
\begin{equation}
\begin{array}{lcl}
a_{\pm}&=&\cosh \gamma\cn\frac{u}{2}\mp\sinh\delta\sn\frac{u}{2}\dn
\frac{u}{2}\\ 
b_{\pm}&=&\cosh \delta\sn\frac{u}{2}\dn\frac{u}{2}\pm\sinh
\gamma\cn\frac{u}{2}\\
c&=&\dn\frac{u}{2}\\
d&=&k\sn\frac{u}{2}\cn\frac{u}{2}
\end{array}
\end{equation}
with
\begin{equation}
\Gamma=\frac{k}{\cosh\left(\gamma+\delta\right)}\;,\quad \quad\quad h=
\tanh\left(\gamma+\delta\right)
\end{equation}
where $k$ is the modulus of the elliptic functions. While for $|h|>1$
\footnote{The spectral parameter here differs from that in \cite{fel}
by a shift of $K$.}
\begin{equation}
\begin{array}{lcl}
a_{\pm}&=&\cosh\gamma\dn\frac{u}{2}\pm k\cosh\delta\sn\frac{u}{2}\cn
\frac{u}{2}\\ 
b_{\pm}&=&\pm\sinh\gamma\dn\frac{u}{2}- k\sinh\delta\sn\frac{u}{2}\cn
\frac{u}{2}\\ 
c&=&\cn\frac{u}{2}\dn\frac{u}{2}\\
d&=&k^{'}\sn\frac{u}{2}
\end{array}\end{equation}
with
\begin{equation}
\Gamma=\frac{k^{'}}{k\sinh\left(\gamma+\delta\right)}\;,\quad\quad\quad
h=\coth\left(\gamma+\delta\right)
\end{equation}
where $k^{'}$ is the complementary modulus of the elliptic functions.

These $R$-matrices satisfy the unitarity and cross-unitarity properties
\begin{equation}
R(u)R(-u)\propto {\bf 1}\;,\quad\quad\quad R(u)^{{\rm t}_1}R(2K-u)^{{\rm t}_2}
\propto {\bf 1}\;.
\end{equation}
So the crossing parameter is $2K$ a half-period of the 
elliptic functions. In general, due to the asymmetry, the $R$-matrices do 
not have crossing symmetry. For the regime $|h|<1$, 
when $\gamma=\delta$, we have
\begin{equation}
R(2K-u)^{{\rm t}_2}=\sigma_1^xR(u)\sigma_1^x\;,
\end{equation}
hence the basis of the ${\bf C}^2$ space can be given an 
interpretation of up-down spins. 
As for the regime $|h|>1$, when $\gamma=0$, we have instead
\begin{equation}
R(2K-u)^{{\rm t}_1}=R(u)\;.
\end{equation}
A natural interpretation of this property can be given if we regard
the basis of the ${\bf C}^2$ space as two distinct degrees of freedom
such as a boson and a fermion.

Note also that for $\gamma=0$, these $R$-matrices are regular in that
\begin{equation}
R_{12}(0)\propto {\cal P}_{12}
\end{equation}
where ${\cal P}_{12}$ is the exchange operator of the spaces $1,2$.

With the bulk $R$-matrix given, it is straight forward to solve for the 
boundary $K$-matrix
using Eq.(\ref{eq:bybe}). We again find that in order for 
solutions to exist, the
bulk parameter $\gamma$ has to vanish, essentially this is due to
the asymmetry of the vertex weights $b_{\pm}$. With this restriction, 
the most general solution is given as follows:
\begin{description}
\item For $|h|<1$,
\begin{equation}
K(u)=\left(\begin{array}{ccc}
G_{+}(u)\cn \frac{u}{2}\dn \frac{u}{2}-G_{-}(u)\sn \frac{u}{2}&2\epsilon\sn 
\frac{u}{2}\cn \frac{u}{2}\dn \frac{u}{2}\vspace{3mm}\\ 
2\sn \frac{u}{2}\cn \frac{u}{2}\dn \frac{u}{2}&G_{+}(u)\cn 
\frac{u}{2}\dn \frac{u}{2}+G_{-}(u)\sn \frac{u}{2}
\end{array}\right)
\end{equation}
where
\[\begin{array}{lll}
G_{+}&=&\alpha_{+}+\frac{\textstyle k\left(\left(1-\epsilon k\cosh\delta\right)
\alpha_{+}+\sinh\delta\alpha_{-}\right)}{\textstyle 
\epsilon\cosh\delta-k}\sn^2\frac{u}{2}\vspace{3mm}\\
G_{-}&=&\alpha_{-}+\frac{\textstyle k\left(\left(1
-\epsilon k\cosh\delta\right)\alpha_{-}
-{k^{'}}^{2}\sinh\delta\alpha_{+}\right)}{\textstyle \epsilon\cosh\delta-k}
\sn^2\frac{u}{2}\end{array}\;.
\]
Here $\alpha_{\pm}$ are free parameters associated with the boundary
that are related by
\begin{equation}
{k^{'}}^{2}\alpha_{+}^{2}+\alpha_{-}^2=2\left(k^{-1}\cosh\delta-
\epsilon\right)\end{equation}
and $\epsilon=\pm 1$. 
\item For $|h|>1$,
\begin{equation}
K(u)=\left(\begin{array}{ccc}
G_{+}(u)\cn \frac{u}{2}+G_{-}(u)\sn \frac{u}{2}\dn \frac{u}{2}&2\epsilon\sn 
\frac{u}{2}\cn \frac{u}{2}\dn \frac{u}{2}\vspace{3mm}\\
2\sn \frac{u}{2}\cn \frac{u}{2}\dn \frac{u}{2}&G_{+}(u)\cn \frac{u}{2}
-G_{-}(u)\sn \frac{u}{2}\dn \frac{u}{2}
\end{array}\right)\label{eq:eswb}
\end{equation}
where
\[G_{\pm}(u)=\alpha_{\pm}-\sn^2 \frac{u}{2}\frac{\alpha_{\pm}\epsilon k\sinh
\delta+ \alpha_{\mp}k^{'}k\cosh \delta}{k^{'}+\epsilon k\sinh\delta}\;.\]

Again $\alpha_{\pm}$ are free parameters associated with the boundary
that are related by
\[\alpha_{+}^{2}-\alpha_{-}^2=2\epsilon\left(1+\frac{\epsilon k\sinh\delta}
{k^{'}}\right)\]
and $\epsilon=\pm 1$. 
\end{description}
Hence for each regime there are two classes of solution distinguished by
$\epsilon=\pm 1$ with one independent parameter. 

Below we shall list a number of properties of these $K$-matrices.

\begin{enumerate}
\item Regularity
\begin{equation}K(0)\propto {\bf 1}\;.\end{equation}
\item Unitarity
\begin{equation}K(u)K(-u)\propto{\bf 1}\;.\end{equation}
\item Corresponding Quantum Spin Chains

With the boundary $K$-matrix and the bulk $R$-matrix, one can
construct an integrable quantum spin chain with fixed boundary
condition using the prescription given in \cite{skly}. The quantity
that generates the spin chain Hamiltonian and other commuting conserved
charges is 
\[ t(u)={\rm tr}\left(K(K-u;{\bf \alpha})T(u)K(u;\tilde{\bf\alpha})T^{-1}(-u)
\right)\]
where $T(u)$ is the usual bulk monodromy matrix constructed out of
the $R$-matrix, and ${\bf \alpha},\tilde{\bf\alpha}$ denote, respectively, 
the parameters $\alpha_{\pm},\tilde{\alpha}_{\pm}$ that are associated with 
the two boundaries of the spin chain. A new feature of the model is that
${\rm tr}\left(K(K;\tilde{\bf \alpha})\right)$ vanishes, as a result we have
\[t(u)=({\rm const.})u{\bf 1}+({\rm const.})u^2{\cal H}+\cdots\;,\]
which is common in models with a supersymmetry.\footnote{The same
property is also observed in \cite{cue} for a special case of this model.}
In the $|h|<1$ regime, we find
\begin{eqnarray}
{\cal H}&\propto& \sqrt{\cosh^2\delta-k^2}\sum_{j=1}^{N-1}\left[
\sqrt{\frac{\cosh\delta+k}{\cosh\delta-k}}\sigma_j^{x}\sigma_{j+1}^{x}
+\sqrt{\frac{\cosh\delta-k}{\cosh\delta+k}}\sigma_j^{y}\sigma_{j+1}^{y}\right.
\nonumber\\
&&\mbox{}\left.-\sqrt{\frac{\sinh\delta+ik^{'}}{\sinh\delta-ik^{'}}}
\sigma_j^{z}-\sqrt{\frac{\sinh\delta-ik^{'}}{\sinh\delta+ik^{'}}}
\sigma_{j+1}^{z}\right]+\left(ik^{'}-
\frac{\tilde{\alpha}_{-}}{\tilde{\alpha}_{+}}\right)\sigma^z_{1}
\nonumber\\
&&\mbox{}+\frac{2}{\tilde{\alpha_{+}}}\left(\sigma_1^{-}
+\epsilon\sigma_1^{+}\right)
+\frac{1-\epsilon k\cosh\delta-ik^{'}\sinh\delta}
{1-\epsilon k\cosh\delta+\frac{\alpha_{-}}{\alpha_{+}}\sinh\delta}
\left(\frac{\alpha_{-}}{\alpha_{+}}-ik^{'}\right)\sigma_N^{z}\nonumber\\
&&\mbox{}+\frac{2\left(k+\epsilon\cosh\delta\right)}
{\alpha_{+}\left(1-\epsilon k\cosh\delta\right)+\alpha_{-}\sinh\delta}
\left(\epsilon\sigma^{-}_{N}+\sigma^{+}_{N}\right)\;.
\end{eqnarray}
While for $|h|>1$, we find
\begin{eqnarray}
{\cal H}&\propto& -\sqrt{k^2\cosh^\delta-1}\sum_{j=1}^{N-1}\left[
\sqrt{\frac{k\sinh\delta-k^{'}}{k\sinh\delta+k^{'}}}\sigma_j^{x}
\sigma_{j+1}^{x}+\sqrt{\frac{k\sinh\delta+k^{'}}{k\sinh\delta-k^{'}}}
\sigma_j^{y}\sigma_{j+1}^{y}\right.\nonumber\\
&&\mbox{}\left.-\sqrt{\frac{k\cosh\delta+1}{k\cosh\delta-1}}\sigma_j^z
-\sqrt{\frac{k\cosh\delta-1}{k\cosh\delta+1}}\sigma_{j+1}^{z}\right]
+\left(\frac{\tilde{\alpha}_{-}}{\tilde{\alpha}_{+}}-1\right)
\sigma_1^{z}\nonumber\\
&&\mbox{}+\frac{2}{\tilde{\alpha}_{+}}\left(\sigma_1^{-}
+\epsilon\sigma_1^{+}\right)+\left(\frac{\alpha_{-}}{\alpha_{+}}
+1\right)\sigma_N^{z}+\frac{2}{\alpha_{+}}\left(\epsilon\sigma^{-}_{N}
+\sigma^{+}_{N}\right)\;.
\end{eqnarray}
\item Quantum Group symmetry

These spin chains include, as a special case, that obtained by \cite{cue}
which has the quantum group $U_{p,q}(gl(1|1))$ ( or $CH_q(2)$ ) symmetry
where $p,q$ are related to $k,\delta$ (see \cite{cue,rit,gom}). 
This can be seen by taking the limit 
$|\alpha_{+}|,|\alpha_{-}|\left(|\tilde{\alpha}_{+}|,
|\tilde{\alpha}_{-}|\right)
\rightarrow\infty$, which, from the
relation satisfied by them, implies that the ratio of them is finite. 
For the regime $|h|<1$, taking $\frac{\tilde{\alpha}_{-}}{\tilde{\alpha}_{+}}
=\frac{\alpha_{-}}{\alpha_{+}}=\pm ik^{'}$ gives the quantum group symmetric
spin chain. While for $|h|>1$, we have to take 
$\frac{\tilde{\alpha}_{-}}{\tilde{\alpha}_{+}}
=-\frac{\alpha_{-}}{\alpha_{+}}=\pm 1$. Note that in this limit the 
$K$-matrices are diagonal.
 
\item Symmetric Limit

It is well known that in the symmetric limit $\delta=0$, ( in addition to
$\gamma=0$ ) the $R$-matrix for the regime $|h|<1$ is a special case
of Baxter's symmetric eight-vertex model. Therefore, in the symmetric 
limit, the $K$-matrix in the $|h|<1$ regime should also be a special
case of that obtained in \cite{hou,inam1}. Recall that there are three
boundary free parameters $\xi,\lambda,\mu$ there, but we have only
one free parameter here. We find that indeed our solution corresponds
to the case where one of the above three parameters vanishes with the
remaining two related. It is intriguing that this way of approaching 
the symmetric free fermion gives rise to $K$ matrices of less boundary 
free parameters.    
\end{enumerate}


\newpage
\begin{thebibliography}{99}
\bibitem{cher} I.V.Cherednik, Teor. Mat. Fiz. {\bf 61} (1984) 35.
\bibitem{skly} E.K.Sklyanin, J. Phys. {\bf A21} (1988) 2375.
\bibitem{oth} L.Mezincescu and R.I.Nepomechie, Int. J. Mod. Phys. 
{\bf A6} (1991) 5231.
\bibitem{zam} S.Ghoshal and A.B.Zamolodchikov, Int. J. Mod. Phys. 
{\bf A9} (1994) 3841.
\bibitem{inam} T.Inami, S.Odake and Y.-Z.Zhang, Phys. Lett. {\bf B359}
(1995) 118.
\bibitem{ssg} C.Ahn and W.M.Koo, ``Exact boundary scattering matrices of
the supersymmetry sine-Gordon theory on a half line", {\tt hep-th/9509056};
Nucl. Phys. {\bf B468} (1996) 461. 
\bibitem{bl} C.Ahn, D.Bernard and A.LeClair, Nucl. Phys. {\bf B346} 
(1990) 409.
\bibitem{karow} M.Karowski and H.J.Thun, Nucl. Phys. {\bf B130} (1977) 295.
\bibitem{ahn} C.Ahn, Nucl. Phys. {\bf B354} (1991) 57;
Prog. Theor. Phys., {\bf 118} (1995) 165.
\bibitem{shan} R.Shankar and E.Witten, Phys. Rev. {\bf D17} (1978) 2134.
\bibitem{fan} C.Fan and F.Y.Wu, Phys. Rev. {\bf B2} (1970) 723.
\bibitem{dev} H.J.de Vega and A.Gonzalez-Ruiz, J. Phys. {\bf A27} (1994) 6129.
\bibitem{warner} N. Warner, Nucl. Phys. {\bf B450} (1995) 663.
\bibitem{ghos} S.Ghoshal, Int. J. Mod. Phys., {\bf A9} (1994) 4801. 
\bibitem{fel} B.U.Felderhof, Physica {\bf 65} (1973) 421; {\em ibid}, 
{\bf 65} (1973) 279, 509. 
\bibitem{baz} V.V.Bazhanov and Yu.G.Stroganov, Teo. Mat. Fiz., {\bf 62} 
(1984) 377;
{\em ibid} {\bf 63} (1984) 291; {\em ibid} {\bf 63} (1984) 417.
\bibitem{cue} R.Cuerno and A.Gonzalez-Ruiz,J. Phys. {\bf A26} (1993) L605. 
\bibitem{rit} H.Hinrichsen and V.Rittenberg, Phys. Lett. {\bf B304} (1993)115;
{\em ibid} {\bf B275} (1992) 350.
\bibitem{gom} R.Cuerno, C.Gomez, E.Lopez-Manzanares and G.Sierra, Phys. Lett.
{\bf 307} (1993) 56.
\bibitem{hou} B.Y.Hou and R.H.Tue, Phys. Lett. {\bf A183} (1993) 169;
\bibitem{inam1} T.Inami and H.Konno, J. Phys. {\bf A27} (1994) L913.
\bibitem{schout} M. Moriconi and K. Schoutens, ``Reflection Matrices
for Integrable $N=1$ Supersymmetric Theories'',
{\tt hep-th/9505219}.
\end{thebibliography}
\end{document}